\documentclass[twocolumn,twoside,slac]{revtex4}
\usepackage{graphicx}
\usepackage{fancyhdr}
\begin{document}

\title{OVAL: the CMS Testing Robot}

%

\author{D. Chamont}
\author{C. Charlot}
\affiliation{ LLR, IN2P3, CNRS, Ecole Polytechnique, France}

\begin{abstract}
Oval is a testing tool which help developers to detect unexpected changes
in the behavior of their software. It is able to automatically compile some
test programs, to prepare on the fly the needed configuration files, to run
the tests within a specified Unix environment, and finally to analyze the
output and check expectations. Oval does not provide utility code to help
writing the tests, therefore it is quite independant of the programming/scripting
language of the software to be tested. It can be seen as a kind of robot which
apply the tests and warn about any unexpected change in the output. Oval was
developed by the LLR laboratory for the needs of the CMS experiment, and it
is now recommended by the CERN LCG project.
\end{abstract}

\maketitle

\thispagestyle{fancy}


\section{Introduction}


New agile methologies \cite{XPROG} put the test programs at the center of the software
development: they should be written before the source code, and they should be executed
whenever the code is modified. This optimistic goal can only be approached if this task
is really easy and as automated as possible. Oval is a tool which tries to fulfill this
need.


Oval \cite{OVAL} was initially developed in the context of CMS \cite{CMS} Software Quality
Improvement to support the high level validation of physics algorithms. We then felt the tool
could also be used for low level unitary tests, and it appears to be actually its first use.


As a CMS tool, Oval was primarily developed for the Unix world, and given a look-and-feel
close to Scram \cite{SCRAM}, the CMS tool for configuration and build. That means it has
a simple command line interface, similar to the CVS one \cite{CVS}, the Oval
configuration files are XML-like \cite{XML}, and the whole is written in PERL.

Oval is mainly a robot which runs the tests for you. It does not provide any helper code,
therefore it is quite independant from your programming or scripting language, but you must
write your own tests. Another key point is that Oval does not analyze some dedicated
log files or output channels, but the standard Unix output. This makes it only usable with
programs which use this way of communication. On the other hand, this allows to scan any
output from any third party software, even when this software has not been especially
instrumented for Oval.


While Oval was appearing useful and usable outside of CMS, it has been progressively
isolated from Scram and extended with many configuration possibilities. Today, it can
be used either by a single programmer with few C++ files and a simple makefile, or by
a large collaboration with complex software processes and tools.


In this paper, we will first browse the main features of Oval: how it must be
configured by the users, what are the main commands and what one will find in the
log files. At last, we will discuss the many ways which can be used
by the administrators so to customize Oval.

\section{Setup of a new test directory}

Given a directory with test programs, one must create a special file
called {\tt "OvalFile"}, where one specifies all the information needed
by Oval about the local tests. The first thing to do is to list the
programs which must be taken into consideration. For example:

\begin{verbatim}
<program name="Clusters.cpp">
<program name="Electrons.cpp"
<program name="EnergyFlow.cpp">
\end{verbatim}

If your directory is dedicated to the test, you will surely list here all the programs of the directory. Then, you can ask for the compilation of the programs by typing {\tt "oval build"}. In this case, Oval does not do much, but simply delegates the compilation to the tool which has been specified by your Oval administrator. Yet, the output of the compilation is stored in a log file for each program. In our example, three files will be created ({\tt "Clusters.log"},
{\tt "Electrons.log"} and {\tt "EnergyFlow.log"}). The content of {\tt "Electrons.log"} could be this:

\begin{verbatim}
[oval build] ===========================
[oval build] make Electrons
[oval build] ===========================

c++ -O2 -o Electrons Electrons.o
  -L/home/chamont/ORCA_5/lib/Linux__2.2
  -L/opt/cms/Releases/ORCA_5/lib/Linux__2.2
  -L/cern/2001/lib
  -lElectronFacilities -lEgammaH4Support
  -lz -lnsl -ldl -lg2c -lm
\end{verbatim}

Some header lines (starting with {\tt "[oval build]"}) recall the
actual compilation command which has been used, and the lines after
are the compilation output (reformatted by Oval so to improve legibility).

\section{Runtime conditions}

Once all the programs compile successfully, one must care about
their runtime conditions. Again, {\tt "OvalFile"} is the place where to
specify this information. Here is the new content we could give to the
configuration file:

\begin{verbatim}
<var name="FEDERATION" value="cmsuf01">
<environment name="pt15">
  <var name="DATASET" value="eg_ele_pt15">
  <program name="Clusters.cpp">
  <program name="Electrons.cpp"
     args="-geo detailed">
</environment>
<environment name="flow">
  <var name="DATASET" value="jm_minbias">
  <program name="EnergyFlow.cpp">
</environment>
\end{verbatim}

The second program, {\tt "Electrons"}, will be executed
with the command-line arguments {\tt "-geo detailed"}.
Also, before running any of the programs, Oval will set the Unix
variable {\tt "FEDERATION"} to the value {\tt "cmsuf01"}. On the contrary,
the value of {\tt "DATASET"} depends on the programs. The first
two programs will see a value of {\tt "eg\_ele\_pt15"}, although the
third one will see a value of {\tt "jm\_minbias"}. The XML tags
are self-explanatory and do not require much more comments.

On top of Unix variables, on can also specify some files to be
created on the fly, before running the programs. We call them
"auxiliary files". It is rarely relevant to specify by this
way the input data (often huge and/or binary), but this
applies well for the configuration files of the software to
be tested, especially if you want to specify different
file contents for the different programs. Here is an example
of such a specification, to be added to our {\tt "OvalFile"}:

\begin{verbatim}
<file name=".orcarc">
  GoPersistent = 1
  MaxEvents = 500
  Random:Seeds = 0 3
</file>
\end{verbatim}

If only a subset of the file depends on the program, one can
specify the common part at the top level of the {\tt "OvalFile"},
and each specific part in the relevant environments. When
Oval prepares an auxiliary, it first searches
for the different parts and concatenates them.

You can now run all your programs with the command {\tt "oval run"}.
You can also run one program, for example {\tt "oval run Electrons"}. When
executing each program, Oval adds a new section to the log file
of the program:

\begin{verbatim}
[oval run] =============================
[oval run] LD_LIBRARY_PATH =
[oval run]   /home/chamont/ORCA_5/lib
[oval run]  :/opt/cms/Releases/ORCA_5/lib
[oval run]  :/opt/Objectivity/5.2.1/lib
[oval run] FEDERATION = cmsuf01
[oval run] DATASET = eg_ele_pt15
[oval run] .orcarc:
[oval run]   GoPersistent = 1
[oval run]   MaxEvents = 500
[oval run]   Random:Seeds = 0 3
[oval run] =============================

                Welcome to COBRA
...
\end{verbatim}

The header lines (starting with {\tt "[oval run]"}) recall the Unix
variables which have been defined and the content of the auxiliary files,
followed by the real output of the program.

\section{Analysis of the programs output}

If the software system to be tested is a little verbose, you
will never get exactly the same output from one software release
to the other, and Oval must be told what to compare, otherwise it
will always warn you about irrelevant differences. As you can guess,
this is done by adding tags to {\tt "OvalFile"}, and this can differ
from one program to the others if you enclose the comparison tags
and the programs in different environments. Let's comment a trivial
example of comparison tags:

\begin{verbatim}
<diffline expr="^OVAL:">
<diffnumber expr="^energy: (.*)$"
            tolerance="5%">
\end{verbatim}

The first tag {\tt "<diffline>"} is the default behavior of Oval: it
specifies to compare all the lines which start with {\tt "OVAL:"}. It is not
enough if you want to scan output from a third party piece of software,
where you cannot insert {\tt "OVAL:"} at the beginning of relevant lines.
Then you can put additional tags {\tt <diffline>}, where the value of
the attribute {\tt expr} is a PERL regular expression describing some
relevant lines.

In the example above, the second tag is more interesting if you want
to track the value of some important physics quantities. This tag defines
some numbers to be checked: the regular expression specifies
which lines to take into consideration, with the parenthesis specifing where
the number within the line is, and the second attribute
specifying how much the value can differ from the reference.

Once you have specified your comparison strategy, you must eventually
modify your programs so to display the expected outputs, and run
them until their outputs can be considered as a good reference. Then,
the command {\tt "oval validate"} will register the current
outputs as the references. Basically, Oval simply copies each file
{\tt "<program>.log"} as {\tt "<program>.ref"}. For example, if we validate
the first two programs with the command {\tt "oval validate Clusters
Electrons"}, the content of the directory will be:

\begin{verbatim}
cms038:ElectronPhoton> ls
  Makefile
  Clusters.cpp
  Clusters.log
  Clusters.ref
  Electrons.cpp
  Electrons.log
  Electrons.ref
  EnergyFlow.cpp
  EnergyFlow.log
  OvalFile
\end{verbatim}

\section{Validation sessions}

At this stage, your test directory is quite ready to be used.
Each time the software to be tested is modified, you can
move to the directory and type the top level command {\tt "oval prod"}.
For each program which has a reference, Oval will
call {\tt "oval build"}, then {\tt "oval run"}, and finally {\tt "oval diff"}
which will perform the comparison between the log file and the ref
file. The output of Oval could look like this:

\begin{verbatim}
cms038:ElectronPhoton> oval prod
  Clusters: build, run, diff.
  Electrons: build, run, diff (DIFFS).
\end{verbatim}

Above, Oval has noticed a difference between the Electrons output and the
reference. If you want more details, you can run separately the command
{\tt "oval diff Electrons"} or look at the new section which has been
added at the end of {\tt "Electrons.log"}:

\begin{verbatim}
[oval diff] ===========================
[oval diff] diff line: /^OVAL:/
[oval diff] diff number: /^energy: (.*)$/ ~5%
[oval diff] ===========================

ref#1452 != log#2053
ref: OVAL: 12 electrons
---
log: OVAL: 11 electrons

ref#1972 !~ log#2592 (>5%)
ref: energy: 29.7275
---
log: energy: 27.4728
\end{verbatim}

In this log file, the header lines recall the comparison tags which have been used,
followed by a description of all the differences, with a format inspired from the
Unix diff command.

\section{Other Features}

Since it has been developed and used for several years, Oval has accumulated
many other features. Let's list the important ones:

\begin{enumerate}

\item{} Oval is not restrained to programs. For complex testing tasks (for
example when databases are involved), one can also run executable scripts or
ready binaries which do not need to be built.

\item{} While a test is getting more complex and is given runtime options,
one could want to run this test several times with different runtime conditions
(commandline arguments, Unix variables, auxiliary files). An {\tt "OvalFile"}
allows several occurrences of the same program. Each occurence will have its
own log file.

\item{} If there are many tests, they should be organized into a hierarchy of
Unix directories. One can insert an {\tt "OvalFile"} at any level: the
specifications will be recursively propagated to the subdirectories.

\item{} The results of a validation session can be mailed to the watchers
specified in the configuration. This is the seed of a future interface
with more elaborated bug tracking tools.

\end{enumerate}


\section{Site Customization}

Oval is potentially collaborating with many other actors: the operating system,
the configuration tool, the build tool, the compiler, the run tool, the bug tracking
system, etc. No two teams are using the same tools and apply the same software process.
This is why it is especially important to make Oval as flexible as possible. So
to demonstrate it, this section focuses on what can be done by a site administrator.
We will begin with the description of the versions and flavors control, then have a
look at the interfaces with the external tools, and finally discuss how to add new
local commands.

\subsection{Versions and Flavors}

All the Oval versions of a site are expected to be installed under a common directory,
which we will call {\tt "OVAL\_DIR"}. When an oval executable is invoked, it first compares
its internal version with the one required by the user, and if it is not the same, it is
able to find the correct executable under {\tt "OVAL\_DIR"} and execute this one instead.
The user can require a specific version by setting the variable {\tt "OVAL\_DIR"}, or thanks
to the tag {\tt "<oval version=...>"} within an {\tt "OvalFile"}. The latter allows to
attach each set of test programs to a given oval version, so that it will
still work when the site administrator will install a new oval version and
make it the default (provided this administrator does not erase the old
oval versions...). This is not only useful, but absolutely necessary for
a site with many users.

On top of the version, the users can also select a given "flavor", thanks to the
variable {\tt "OVAL\_FLAVOR"}. Depending on the flavor, oval will use different default
configurations, so to fill the needs of different user groups. For example, if the value
of {\tt "OVAL\_FLAVOR"} is {\tt "salty"}, oval will take its default values from
{\tt "OVAL\_DIR/site/OvalFile.salty"}. The definition of the flavors and the corresponding
configurations is up to the site administrators.

Actually, whenever Oval needs a file which can be customized (for example {\tt "OvalFile"}),
it first scans for this file in the {\tt "OVAL\_DIR/site"} directory, where it will select in
priority the most specialized file it can found, for a given version and flavor (for example
{\tt "OVAL\_DIR/site/version/flavor/OvalFile"} or
{\tt "OVAL\_DIR/site/OvalFile.version.flavor"}).
This lets the site administrators provide default configurations for any
combination of version and flavor, if they need to.

These site files are typically used for the definition of instructions which
are platform-dependant. They can also be used to enforce options on the
commands. Here is an example:

\begin{verbatim}
<options command="expr" value="-v">
<config name="mail instruction"
        value="/bin/mail -s %">
<config name="custom url"
        value="http://www.site.fr/oval">
\end{verbatim}

\subsection{Access to the External Tools}

The use of external tools, such as scram or make, is done through
perl wrappers. The directory {\tt "OVAL\_DIR/OVAL\_VERSION/share/Interfaces"}
contains the wrappers delivered with Oval. Since the features of the various tools
are somehow fuzzy and overlapping, we made the list of all the functions that
the external tools should implement, and we splitted them into few groups, that we
will call "interfaces". There is a global map which associates a concrete tool to
each interface. Whenever Oval needs a given external function, which is part of
a given interface, Oval looks which tool is attached to the interface,
and invokes the function from this tool. Currently, we have defined two
interfaces: {\tt "build"} which is attached by default to the
tool {\tt "make"}, and {\tt "run"} which is attached by default to the
tool {\tt "oval"} (the pseudo tool {\tt "oval"} provides a default implementation
for all the defined interfaces). If the users of a site are using
Scram, the administrator can declare it in {\tt "OVAL\_DIR/site/OvalFile"}:

\begin{verbatim}
<config name="build tool" value="scram">
<config name="run tool" value="scram">
\end{verbatim}

As we described above, this configuration can depend on the version and/or
the flavor. If an administrator want to use a tool unknown to Oval (for which no
wrapper is provided), he can copy one of the provided wrapper from
{\tt "OVAL\_DIR/OVAL\_VERSION/share/Interfaces"} to
{\tt "OVAL\_DIR/site/Interfaces"} and modify it appropriately. One can
also reimplement an existing wrapper: the site wrapper will hide
the default Oval one. This can also depend on the version and/or flavor.
For example, a tool wrapper called {\tt "OVAL\_DIR/site/salty/Interfaces/cmt.pm"}
would only be visible to the users who set {\tt "OVAL\_FLAVOR"} to
{\tt "salty"}, and the tool wrapper {\tt "OVAL\_DIR/site/3\_0\_0/Interfaces/make.pm"}
will hide the default one only for the users who set {\tt "OVAL\_VERSION"} to
{\tt "3\_0\_0"}.

\subsection{Commands}

When a need is so specific that the Oval team can hardly include it in
its standard packaging, the site administrators can provide a specific
site command. If their file respects the Oval commands requirements, and
is placed in a directory called {\tt "Commands"} ( for example
{\tt "OVAL\_DIR/site/Commands/mycom.pm"} ), the new command
will be detected by Oval and made available to the users. As one
can guess, this requires a rather deep understanding of Oval
internals.

Even more "nasty", the administrators can provide their own
implementation of the existing commands.

\section{Conclusion}

As hopefully demonstrated, Oval can be used for simple projects as well
as configured for large collaborations which use other external tools and a strict
software process. The only aspect which could prove a little painful is the design
of the regular expressions.

Any piece of software, written in any language, can be automatically tested
all along its life (provided it is talking to the output channel). Unhapilly,
the programmers still have to write the tests and maintain the reference outputs,
but we are not convinced any tool will ever avoid this.

The CMS collaboration currently uses Oval for its regression tests,
and the CERN LCG SPI group now recommends the use of Oval together with
CppUnit \cite{CPPUNIT} which provide some C++ helper code. Oval is not only useful
for the punctual validation of software releases: in the spirit of agile methods,
it can be used daily so to detect the regressions as soon as they appear.

For the near future, the main task will be to add a tutorial and to react to the
users feedback. We also want to better document the internals, use the
CERN LCG savannah server and open the development of Oval. At last, we
plan to redesign the output and log files of Oval, so to ease the life of
librarians who handle huge hierarchies of tests, and enable to trace
the status of all these tests.


%



\end{document}